\documentclass[a4paper,10pt]{article}
\newcommand{\dfr}{d\raise0.3ex\hbox{\kern-0.6ex\char"013 }} 
\newcommand{\ddfr}{{ }^{-}\!\!\!\!\!\dfr} 
\newcommand{\intfr}{-\!\!\!\!\!\!\int} 
\begin{document}
%
%
\def\ov{\over}
\def\l{\left}
\def\r{\right}
\def\be{\begin{equation}}
\def\ee{\end{equation}}

\title{Quantum-classical transition in Scale Relativity}
\author{{\bf Marie-No\"elle C\'el\'erier$^1$ and Laurent Nottale$^2$} 
\\
LUTH, CNRS, Observatoire de Paris-Meudon, \\
5 place Jules Janssen, 92195 Meudon Cedex, France \\
e-mail: $^1$ marie-noelle.celerier@obspm.fr \\
$^2$ laurent.nottale@obspm.fr \\
Published: J. Phys. A: Math. Gen. {\bf 37} (2004) 931-955 \\
Copyright: IOP Publishing Ltd}

\date{\today} 
\maketitle

\begin{abstract}
The theory of scale relativity provides a new insight into the origin 
of fundamental laws in physics. Its application to microphysics allows us to recover quantum mechanics as mechanics on a non-differentiable 
(fractal) spacetime. The Schr\"odinger and Klein-Gordon equations 
are demonstrated as geodesic equations in this framework. A development of the intrinsic properties of this theory, using the mathematical tool of Hamilton's bi-quaternions, leads us to 
a derivation of the Dirac equation within the scale-relativity paradigm. The complex form of the wavefunction in the Schr\"odinger and Klein-Gordon equations follows from the non-differentiability of the geometry, since it involves a breaking of the invariance under the reflection symmetry on the (proper) time differential element ($ds \leftrightarrow - ds$). This mechanism is generalized for obtaining the bi-quaternionic nature of the Dirac spinor by adding a further symmetry breaking due to non-differentiability, namely the differential coordinate reflection symmetry ($dx^{\mu} \leftrightarrow - dx^{\mu}$) and by requiring invariance under parity and time inversion. The Pauli equation is recovered as a non-relativistic-motion approximation of the Dirac equation.
\end{abstract}


\section{Introduction}
\label{s:intro}

The theory of scale relativity consists of generalizing to scale transformations the principle of relativity, which has been applied 
by Einstein to motion laws. It is based on the giving up of the assumption of spacetime coordinate differentiability, which is usually retained as an implicit hypothesis in current physics. Even though this hypothesis can be considered as roughly valid in the classical domain, it is clearly broken by the quantum mechanical behavior. It has indeed been pointed out by Feynman (see, e.g., \cite{FH65}) that the typical paths of quantum mechanics are continuous but non-differentiable. 

In the present paper, after a reminder about the Schr\"odinger and 
Klein-Gordon equations, we apply the scale-relativistic approach to the understanding of the nature of bi-spinors and of the Dirac 
equation in a spacetime representation. A step in this direction has been made by Gaveau {\it et al} \cite{GJ84} who have generalized Nelson's stochastic mechanics \cite{EN66} to the relativistic-motion case in (1+1) dimensions. However, an analytic continuation was needed to obtain the Dirac equation in such a framework, and, furthermore, stochastic mechanics is now known to be in contradiction with standard quantum mechanics as concerns multitime correlations in repeated measurements \cite{WL93}.

In the framework of a fractal spacetime viewed as a geometric analogue of quantum mechanics \cite{GO83,NS84}, Ord has developed a generalized version of the Feynman chessboard model which allows him to recover the Dirac equation in (3+1) dimensions, without analytical continuation \cite{GO92,MO92,OM93,GO96}. We develop in the present paper an approach involving also a fractal spacetime, but in a different way: namely the fractality of spacetime is derived from the giving up of its differentiability, it is constrained by the principle of scale relativity and the Dirac equation is derived as an integral of the geodesic equation \cite{CN03}. This is not a stochastic approach (the use of statistically defined quantities is derived and we do not use Fokker-Planck equations), so that it does not come under the contradictions encountered by stochastic mechanics. 

One of the fundamental reasons for jumping to a scale-relativistic description of nature is to increase the generality of the first principles retained to construct the theory. From a principle known as Occam's razor, the laws and structures of nature are the most general and simple laws and structures that are physically possible. It culminates nowadays in Einstein's ``general'' relativity and his 
description of gravitation as a manifestation of the Riemannian geometry of spacetime, obtained by giving up the Euclidean flatness hypothesis. However, Einstein's principle of relativity is not yet fully general, since it applies to coordinate transformations that are continuous and, at least two times, differentiable. The aim of the theory of scale relativity is to bring to light laws and structures that would be the manifestation of more general transformations, namely continuous ones, either differentiable or not. The standard ``general relativistic'' theory will be recovered as the special differentiable case, while new effects would be expected from the 
non-differentiable part. 

In the present work, we focus our attention on the microphysical scale of motion in a non-differentiable spacetime, in the framework of 
``Galilean scale relativity" \cite{LN93}, i.e. we consider fractal power law dilations with a constant fractal dimension. Therefore, we recover, from the first principles of this theory, (i) the four main evolution equations of standard quantum physics as geodesic equations on a fractal space/spacetime, namely the Schr\"odinger, (free) Klein-Gordon, Pauli and (free) Dirac equations, and (ii) the nature of the wavefunctions that appear in these equations, which are, respectively, complex, spinor and bi-spinor quantities. Indeed, as we shall see, the non-differentiability has two main consequences: (i) the fractality of the geometry that generates the purely quantum terms in the equations and (ii) discrete symmetry breakings at the infinitesimal level that lead to successive algebra doublings in the description of physical quantities.

The derivation of the Schr\"odinger equation is given in Sec.~\ref{s:schro}, where we actualize the former works dealing with this issue, proposing a more accurate interpretation of the nature 
of the transition from the non-differentiable (fractal quantum scales) to the differentiable (classical scales) domain, and carefully justifying the different key choices made at each main step of the reasoning. An analogous updating is proposed in Sec.~\ref{s:ckgeq}, for the free Klein-Gordon equation acting on a complex wavefunction. 
We obtain, in Sec.~\ref{s:kgeq}, the bi-quaternionic form of the same 
Klein-Gordon equation, from which the Dirac equation then the Pauli equation naturally proceed, as shown in Secs.~\ref{s:dieq} and \ref{s:paeq}. The appendix is devoted to the presentation and justification of the mathematical tools we use. 

\section{The Schr\"odinger equation revisited}
\label{s:schro}

The Schr\"odinger equation is derived, in the framework of scale 
relativity, as a geodesic equation in a fractal three-space (see, e.g., \cite{LN93,LN96A,LN97A}). In the present section, we update and give a more precise meaning to some of the most subtle issues we are dealing with.

\subsection{Transition from non-differentiability (fractal scales) to  
differentiability (classical scales)}
\label{ss:fracsp}

The aim of the present subsection is to identify the consequences of the giving up of the coordinate differentiability. We develop here the 
formalism relevant for the derivation of the non-relativistic-motion Schr\"odinger equation, valid in a fractal three-space with the time $t$ as a curvilinear parameter. 

Strictly, the non-differentiability of the coordinates means that the 
velocity
\be
V = {dX\over dt}= \lim_{dt \rightarrow 0} \frac{X(t+dt) - X(t)}{dt}
\label{eq.15}
\ee
is undefined. This implies that, when $dt$ tends to zero, either the ratio $dX/dt$ tends to infinity, or it fluctuates without reaching any limit. This problem is overcome in the scale relativity framework thanks to the following fundamental theorem:{\it a continuous and non-differentiable (or almost nowhere differentiable) curve is 
explicitly scale dependent, and its length tends to infinity when the scale interval tends to zero} \cite{LN93,LN96A,BC00}. In other words, a continuous and non-differentiable space is fractal, in the general meaning given by Mandelbrot to this concept \cite{BM82}. The velocity $V(t)$, though it is undefined in the standard way, can now be defined in a new way as a ``fractal function" \cite{LN93}, $V(t,\delta t)$, which is explicitly dependent on the scale interval $\delta t$. As recalled in previous works \cite{LN93,LN89,LN92}, these explicit scale variables can be defined only in a relative way (only their ratio has a physical meaning) so that the description of the scale space in which they lie is expected to come under a theory of scale relativity.

In the case of a constant fractal dimension $D$, the relation between the position resolution interval and the time resolution interval is given by \cite{LN93,BM82}
\be
\delta X \approx \delta t ^{1/D}.
\label{eq.16}
\ee

The advantage of this method is that, for any given value of the resolution, $\delta t$, differentiability in $t$ is recovered, which allows us to use the differential calculus, even when dealing with non-differentiability. However, the physical and the mathematical descriptions are not always coincident. Once $\delta t$ is given, one can write mathematical differential equations involving $\partial / \partial t$, make $\partial t \rightarrow 0$, then solve for it and determine $V(t, \delta t)$. Actually, this is a purely mathematical 
description with no physical counterpart, since the very consideration of an interval $dt < \delta t$, as occuring in an actual measurement, changes the function $V$ (such a behaviour, described by Heisenberg's uncertainty relations, is experimentally verified for any quantum system). However, making the particular choice $dt = \delta t$ induces 
a peculiar subspace of description where the physics and the mathematics coincide. We work, for the Schr\"odinger and Klein-Gordon 
equations, with such an identification of the time differential and of the new time resolution variable, but are led to give it up in  Secs.~\ref{s:kgeq} and ~\ref{s:dieq}. 

The scale dependence of the velocity suggests that we complete the 
standard equations of physics by new differential equations of scale. 
Writing the simplest possible equation for the variation of the velocity, $V(t, dt)$, in terms of the new scale variable $dt$, as a first-order differential equation, $\partial V/\partial \ln dt=\beta(V)$, then Taylor expanding it, using the fact that $V<1$ (in 
relativistic-motion units $c=1$), we obtain the solution \cite{LN97A}
\be
V = v + w = v \; \left[1 + \zeta \left(\frac{\tau}{dt}\right)^{1-1/D}\right].
\label{eq.17}
\ee
The velocity is now made of two independent terms of different order, since $v/w \approx dt^{1-1/D}$ is an infinitesimal. We call $v$ the ``classical part'' of the velocity (see below the definition of the 
classical part operator $\overline{{\cal C} \ell}$) and $w$ its explicitly scale-dependent ``fractal part''.  The transition scale $\tau$ and the scaled fluctuation $\zeta$ are chosen such that 
$\overline{{\cal C} \ell}\langle\zeta\rangle = 0$ and 
$\overline{{\cal C} \ell}\langle\zeta^2\rangle = 1$. 

We recognize here the combination of typical fractal behaviour with 
fractal dimension $D$ and of a breaking of the scale symmetry at the scale transition $\tau$. In what follows, $\tau$ will be identified with the Einstein-de Broglie scale of the system ($\tau=\hbar/E$), since $V \approx v$ when $dt \gg \tau$ (classical behaviour), and $V \approx w$ when $dt \ll \tau$ (fractal behaviour). Recalling that $D = 2$ plays the role of a critical dimension \cite{LN93,LN96A}, we recover Feynman's result in the asymptotic scaling domain, $w \propto 
(dt/\tau)^{-1/2}$ \cite{FH65}, which allows us to identify the fractal domain with the quantum one. In this paper, we consider only the case of fractal dimension $2$. 

The above description strictly applies to an individual fractal trajectory. Now, one of the geometric consequences of the non-differentiability and of the subsequent fractal character of space/spacetime itself (not only of the trajectories) is that there is an infinity of fractal geodesics relating any couple of points of this space/spacetime \cite{LN93,LN96A}. We therefore suggest \cite{LN89} that the description of a quantum mechanical particle could be reduced to the geometric properties of the set of fractal geodesics that 
corresponds to a given state of this ``particle''. In such an interpretation, the ``particle'' is not identified with a point mass which would follow the geodesics, but we expect its internal properties, mass, spin and charges, to be defined as geometric properties of the fractal geodesics themselves. As a consequence, any measurement is interpreted as a sorting out (or selection) of the 
geodesics by the measuring device \cite{LN93,LN89}. 

The transition scale appearing in Eq.~(\ref{eq.17}) yields two 
distinct types of behaviour of the system (particle) depending on the resolution at which it is considered. Equation (\ref{eq.17}) multiplied by $dt$ gives the elementary displacement, $dX$, of the system as a sum of two terms
\be
dX = dx + d\xi,
\label{eq.18}
\ee
$d\xi$ representing the ``fractal part'' and $dx$, the ``classical part'', defined as
\be
dx = v \; dt,
\label{eq.19}
\ee
\be
d\xi=a \sqrt{2 \cal{D}} (dt^{2})^{1/2D},
\label{eq.20}
\ee
which becomes, for $D=2$,
\be
d\xi=a \sqrt{2 \cal{D}} dt^{1/2},
\label{eq.20bis}
\ee
with $2{\cal D}=\tau _0=\tau v^2$ and $\overline{{\cal C} \ell}\langle a \rangle =0, \overline{{\cal C} \ell}\langle a^2 \rangle =1$. The scale $2{\cal D}=\tau _0$ will be subsequently identified with the Compton scale, $\hbar / mc$, i.e. it gives the mass of the particle up to fundamental constants (see the argument in what follows). We note, from Eqs.~(\ref{eq.18}) to (\ref{eq.20bis}), that $dx$ scales as $dt$, while $d\xi$ scales as $dt^{1\over 2}$. Therefore, the behaviour of the system is dominated by the $d\xi$ term in the non-differentiable ``fractal'' domain (below the transition scale), and by the $dx$ one in the differentiable ``classical'' domain (above the transition scale). 

Now, the Schr\"odinger, Pauli, Klein-Gordon and Dirac equations give results applying to measurements performed on quantum objects, but achieved with classical devices, on the classical/quantum interface. The microphysical scale at which a physical system is considered 
induces the sorting out of a bundle of geodesics, corresponding to the scale of the system, while the measurement process implies a smoothing 
out of the geodesic bundle coupled to a transition from the ``fractal'' to the ``classical'' domain. We therefore define an operator $\overline{{\cal C} \ell}\langle\quad\rangle$, which we apply to the fractal variables or functions each time we are drawn to the ``classical'' domain where the $dt$ behaviour dominates. The effect of $\overline{{\cal C} \ell}$ is to extract, from the fractal variables or functions to which it is applied, the ``classical part'', i.e. the part scaling as $dt$. This justifies our above writting 
$\overline{{\cal C} \ell}\langle\zeta\rangle = 0$ and 
$\overline{{\cal C} \ell}\langle\zeta^2\rangle = 1$, as it is straightforward, from Eq.~(\ref{eq.17}), that the terms involving 
$\zeta$ scale as $dt^{1/2}$ and those involving $\zeta^2$ scale as $dt$. This also led us to state, for the $a$ dimensionless coefficient 
in Eq.~(\ref{eq.20bis}), $\overline{{\cal C} \ell}\langle a\rangle=0$ and $\overline{{\cal C} \ell}\langle a^2\rangle=1$. Note the improvement of our new definition in terms of the classical 
part with respect to the previous interpretation in terms of an averaging process \cite{LN93}. 

\subsection{Differential-time symmetry breaking}
\label{ss:difftsb}

Another consequence of the non-differentiable nature of space (spacetime) is the breaking of local differential (proper) time reflection invariance. The derivative with respect to the time $t$ of a differentiable function $f$ can be written twofold
\be
\frac{df}{dt} = \lim_{dt \rightarrow 0}\frac{f(t+dt) - f(t)}{dt} = 
\lim_{dt \rightarrow 0}\frac{f(t) - f(t-dt)}{dt} \; .
\label{eq.21}
\ee

The two definitions are equivalent in the differentiable case. In the 
non-differentiable situation, both definitions fail, since the limits are no longer defined. In the framework of scale relativity, the physics is related to the behaviour of the function during the ``zoom'' operation on the time resolution $\delta t$, here identified with the differential element $dt$, which is now considered as an independent variable. Two functions $f'_+$ and $f'_-$ are therefore defined as explicit functions of the two variables $t$ and $dt$
\be
f'_+(t,dt) = \frac{f(t+dt,dt)-f(t,dt)}{dt} \; ,
\label{eq.22}
\ee
\be
f'_-(t,dt) = \frac{f(t,dt)-f(t-dt,dt)}{dt} \; .
\label{eq.23}
\ee

We pass from one to the other by the transformation $dt \leftrightarrow -dt$ (local differential-time reflection invariance), 
which was an implicit discrete symmetry of differentiable physics. The 
non-differentiable geometry implies that this symmetry is now broken.
When applied to the space coordinates, these definitions yield two velocities that are fractal functions of the resolution, $V_+[x(t),t,dt]$ and $V_-[x(t),t,dt]$. In order to go back to the 
``classical'' domain and derive the ``classical'' velocities appearing in Eq.~(\ref{eq.19}), we smooth out each fractal geodesic in the bundle selected by the zooming process with balls of radius larger than $\tau$. This amounts to carrying out a transition from the non-differentiable to the differentiable domain and leads to define two ``classical'' velocity fields now resolution independent: 
$V_+[x(t),t,dt>\tau] = \overline{{\cal C} \ell}\langle V_+[x(t),t,dt]\rangle=v_+[x(t),t]$ and $V_-[x(t),t,dt>\tau] = \overline{{\cal C} \ell}\langle V_-[x(t),t,dt]\rangle=v_-[x(t),t]$. After the transition, there is no reason for these two velocities to be equal. While, in standard mechanics, the concept of velocity was one-valued, we introduce two velocities instead of one, even when going back to the ``classical'' domain. This two-valuedness of the velocity vector finds its origin in a breaking of the discrete time reflection invariance symmetry ($dt \leftrightarrow -dt$), which is itself a mathematical consequence of non-differentiability. Therefore, if one reverses the sign of the time differential element, $v _{+}$ becomes $v _{- }$. A natural solution to this problem is to consider both $(dt > 0)$ and  $(dt < 0) $ processes on the same footing, and to combine them in a unique twin process in terms of which the invariance by reflection is recovered. The information needed to describe the system is doubled with respect to the standard description. A simple and natural way to account for this doubling is to use complex numbers and the complex product. It is the origin of the complex nature of the wavefunction of quantum mechanics (see the appendix). 

\subsection{Covariant derivative operator}
\label{ss:scderiv}

Finally, we describe, in the scaling domain, the elementary displacements for both processes, $dX_\pm$, as the sum of a classical part, $dx_\pm = v_\pm\;dt$, and a fluctuation about this classical part, $d\xi_\pm$, which is of zero classical part, $\overline{{\cal C} \ell}\langle d\xi_\pm\rangle = 0$.
\begin{eqnarray} 
  dX_+(t) = v_+\;dt + d\xi_+(t) , \nonumber \\
  dX_-(t) = v_-\;dt + d\xi_-(t) .
\label{eq.24}
\end{eqnarray}

Two classical derivatives, $d/dt_+$ and $d/dt_-$, are defined, using the classical part extraction procedure. Applied to the position 
vector, $x$, they yield two classical velocities
\be
\frac{d}{dt_+}x(t)=v_+ \; , \qquad \frac{d}{dt_-}x(t) = v_- \; .
\label{eq.25}
\ee

As regards the fluctuations, the generalization to three dimensions of the fractal behaviour of Eq.~(\ref{eq.20}) is written (for $D=2$)
\be
\overline{{\cal C} \ell}\langle d\xi_{\pm i}\;d\xi_{\pm j}\rangle = \pm 2 \; 
{\cal D} \; \delta _{ij} \; dt \qquad i,j=x,y,z,
\label{eq.26}
\ee
The classical part of every crossed product $d\xi_{\pm i}\;d\xi_{\pm j}$, with $i\neq j$, is null. This is due to the fact that, even if each term in the product scales as $dt^{1/2}$, each of them behaves as an independent fractal fluctuation around its own classical part. Therefore, when we smooth out the geodesic bundle during the transition from the fractal to the classical domain, we apply a process which is mathematically (not physically) equivalent to a stochastic ``Wiener'' process, and also more general, since we do not need any Gaussian distribution assumption. Thus, we can apply to the classical part of the $d\xi(t)$ product the property of the product of two independent stochastic variables: i.e. the classical part of the product is the product of the classical parts, and therefore here zero. 

To recover local differential time reversibility in terms of a new complex process \cite{LN93}, we combine the two derivatives to 
obtain a complex derivative operator (see the appendix for a more detailed justification)
\be
\frac{\dfr}{dt} = {1\over 2} \left( \frac{d}{dt_+} + \frac{d}{dt_-} \right) 
- {i\over 2} \left(\frac{d}{dt_+} - \frac{d}{dt_-}\right) \; .
\label{eq.27}
\ee

Applying this operator to the position vector yields a complex velocity 
\be
{\cal V} = \frac{\dfr}{dt} x(t) = V -i U = \frac{v_+ + v_-}{2} - i 
\;\frac{v_+ - v_-}{2} \; .
\label{eq.28}
\ee

The minus sign in front of the imaginary term is chosen here in order to finally obtain the Schr\"odinger equation in terms of $\psi$. The reverse choice would give the Schr\"odinger equation for the complex conjugate of the wavefunction $\psi^{\dag}$, and would be therefore physically equivalent. The real part, $V$, of the complex velocity, ${\cal V}$, represents the standard classical velocity, while its imaginary part, $U$, is a new quantity arising from non-differentiability. At the usual classical limit, $v_+ = v_- = v$, 
so that $V=v$ and $U = 0$. 

Contrary to what happens in the differentiable case, the total derivative with respect to time of a fractal function $f(x(t),t)$ of integer fractal dimension contains finite terms up to higher order \cite{AE05} 
\be
{df\over {dt}}= {\partial f \over {\partial t}} + {\partial f \over 
{\partial x_i}}{dX_i \over {dt}} + {1\over 2}{\partial^2 f \over {\partial 
x_i \partial x_j}}{dX_idX_j\over {dt}} + {1\over 6}{\partial^3 f \over 
{\partial x_i \partial x_j \partial x_k}}{dX_idX_jdX_k\over {dt}} + ...
\label{eq.29}
\ee

In our case, a finite contribution only proceeds from terms of $D$-order, while lesser-order terms yield an infinite contribution and higher-order ones are negligible. For a fractal dimension $D=2$, the total derivative is written
\be
{df\over {dt}} = \frac{\partial f}{\partial t} + \nabla f . {dX\over {dt}} + 
\frac{1}{2} \frac{\partial ^2 f}{\partial x_i \partial x_j} {dX_i dX_j 
\over {dt}} \; .
\label{eq.30}
\ee

Usually the term $dX_i dX_j /dt$ is infinitesimal, but here its classical part reduces to $\overline{{\cal C} \ell}\langle d \xi_i\; d\xi_j\rangle /dt$, which is now finite. Therefore, thanks to 
Eq.~(\ref{eq.26}), the last term of the classical part of Eq.~(\ref{eq.30}) amounts to a Laplacian, and we obtain  
\be
{df\over {dt_\pm}} = \left({\partial \over {\partial t}} + v_\pm . \nabla 
\pm {\cal D} \Delta\right) f \; .
\label{eq.31}
\ee

Substituting Eqs.~(\ref{eq.31}) into Eq.~(\ref{eq.27}), we finally 
get the expression for the complex time derivative operator \cite{LN93}
\be
\frac{\dfr}{dt} = \frac{\partial}{\partial t}Ñ + {\cal V}. \nabla - 
i {\cal D} \Delta \; .
\label{eq.32}
\ee

The passage from standard classical (everywhere differentiable) mechanics to the new non-differentiable theory is then implemented by
replacing the standard time derivative $d/dt$ by the new complex operator $\dfr/dt$ \cite{LN93}. In other words, this means that $\dfr/dt$ plays the role of a ``covariant derivative operator'', i.e. a tool that implements the form invariance of the equations. Note that in this replacement, one should be cautious and take into account the fact that this operator is a linear combination of first- and second-order derivatives, in particular when applying it to products and composed functions (see \cite{JCP99A} and the appendix about the Leibniz rule for this operator).

\subsection{Covariant mechanics induced by scale laws}
\label{ss:scmech}

Let us now recall how one generalizes the standard classical mechanics using this covariance. We assume that the classical part of the system can be characterized by a Lagrange function ${\cal L} (x, {\cal V}, t)$, from which an action ${\cal S}$ is defined.
\be
{\cal S} = \int_{t_1}^{t_2} {\cal L} (x, {\cal V}, t) dt.
\label{eq.33}
\ee

The Lagrange function and the action are now complex and are obtained from the classical Lagrange function $L (x, dx/dt, t)$ and the classical action $S$ by replacing $d/dt$ by $\dfr/dt$. The stationary action principle applied on this complex action yields generalized Euler-Lagrange equations (see the appendix),
\be
\frac{\dfr}{dt} \frac{\partial {\cal L}}{\partial {\cal V}_i} = 
\frac{\partial {\cal L}}{\partial x_i} \; ,
\label{eq.34}
\ee
which are the equations one would obtain from applying the covariant derivative operator ($d/dt \rightarrow \dfr/dt$) to the standard 
Euler-Lagrange equations. This demonstrates the self-consistency of the approach and vindicates the use of complex numbers. Other fundamental results of standard classical mechanics are also 
generalized in the same way. In particular, assuming homogeneity of space in the mean leads one to define a generalized complex momentum given by
\be
{\cal P} = \frac{\partial {\cal L}}{\partial {\cal V}}.
\label{eq.35}
\ee

If we now consider the action as a functional of the upper limit of 
integration in Eq.~(\ref{eq.33}), the variation of the action from one 
trajectory to another nearby one yields a generalization of another 
well-known relation of mechanics \cite{LL69}
\begin{equation}
{\cal P} = \nabla {\cal S}.
\label{eq.36}
\end{equation}

\subsection{Generalized Newton-Schr\"odinger equation}
\label{ss:gnscheq}

Let us now consider the general case when the structuring external field is a scalar potential, $\Phi$. The Lagrange function of a closed system, $L=\frac{1}{2}mv^2 -\Phi$, is generalized, in the classical domain, as ${\cal L} (x,{\cal V},t)=\frac{1}{2}m{\cal V}^2-\Phi$. The Euler-Lagrange equations keep the form of Newton's equation of dynamics
\be
m \frac{\dfr}{dt} {\cal V}= - \nabla \Phi ,
\label{eq.37}
\ee
which is now written in terms of complex variables and complex operators. 

In the case when there is no external field, the covariance is explicit, since Eq.~(\ref{eq.37}) takes the form of the equation of inertial motion
\be
\dfr {\cal V} /dt = 0.
\label{eq.37bis}
\ee

In both cases, the complex momentum ${\cal P}$ reads
\be
{\cal P} = m {\cal V} ,
\label{eq.38}
\ee
so that, from Eq.~(\ref{eq.36}), the complex velocity ${\cal V}$ appears as a gradient, namely the gradient of the complex action,
\be
{\cal V} = \nabla {\cal S}/ m.
\label{eq.39}
\ee

We now introduce a complex wavefunction $\psi$ which is nothing but another expression for the complex action ${\cal S}$,
\be
\psi = e^{i{\cal S}/{\cal S}_{0}}.
\label{eq.40}
\end{equation}

The factor ${\cal S}_{0}$ has the dimension of an action and must be 
introduced at least for dimensional reasons. The $\psi$ function is therefore related to the complex velocity appearing in Eq.~(\ref{eq.39}) as follows
\be
{\cal V} = - i \, \frac{{\cal S}_{0}}{m} \, \nabla (\ln \psi).
\label{eq.41}
\ee

The fundamental equation of dynamics (\ref{eq.37}) in terms of the new quantity $\psi$ takes the form
\be
i {\cal S}_{0} \frac{\dfr}{dt}(\nabla \ln \psi) = \nabla \Phi.
\label{eq.42}
\ee

Now ${\dfr}$ and $\nabla$ do not commute. However, there is a particular choice of the arbitrary constant ${\cal S}_{0}$ for which ${\dfr}(\nabla \ln \psi)/dt$ is nevertheless a gradient. 

Replacing $\dfr/dt$ by its expression, given by Eq.~(\ref{eq.32}), and replacing ${\cal V}$ by its expression in Eq.~(\ref{eq.41}), we obtain 
\be
\nabla   \Phi  =   i {\cal S}_{0} \left[ \frac{\partial }{\partial t} \nabla   
\ln\psi   - i \left\{  \frac{{\cal S}_{0}}{m} (\nabla   \ln\psi  . \nabla   )
(\nabla   \ln\psi ) + {\cal D} \Delta (\nabla   \ln\psi )\right\}\right] .
\label{eq.44}
\ee

Using the remarkable identities
\be
(\nabla \ln f)^{2} + \Delta \ln f =\frac{\Delta f}{f} \; ,
\label{eq.45}
\ee
\be
\nabla \Delta =\Delta \nabla , 
\label{eq.48}
\ee
\be
\nabla (\nabla f)^{2}=2 (\nabla f . \nabla) (\nabla f) ,
\label{eq.49}
\end{equation}
we obtain
\be
\nabla\left(\frac{\Delta \psi}{\psi}\right)= 2 (\nabla \ln\psi . \nabla )
(\nabla \ln \psi )  + \Delta (\nabla \ln \psi).
\label{eq.50}
\ee

We recognize, on the right-hand side of this equation, the two terms of Eq.~(\ref{eq.44}), respectively, in factor of ${\cal S}_{0}/{m}$ and ${\cal D}$. Therefore, the choice 
\be
{\cal S}_{0}=2 m {\cal D}
\label{eq.51}
\ee
allows us to simplify this right-hand side, which becomes a gradient. The wavefunction in Eq.~(\ref{eq.40}) is therefore defined as
\be
\psi = e^{i{\cal S}/2m{\cal D}},
\label{eq.52}
\ee
and is a solution of the fundamental equation of dynamics, Eq.~({\ref{eq.37}), written as
\be
\frac{\dfr}{dt} {\cal V} = -2 {\cal D} \nabla \left\{i \frac{\partial}
{\partial t} \ln \psi + {\cal D} \frac{\Delta \psi}{\psi}\right\} = 
-\nabla \Phi / m.
\label{eq.53}
\ee

This equation can now be integrated in a universal way which yields
\be
{\cal D}^2 \Delta \psi + i {\cal D} \frac{\partial}{\partial t} \psi - 
\frac{\Phi}{2m}\psi = 0,
\label{eq.54}
\ee
up to an arbitrary phase factor which may be set to zero by a suitable 
choice of the phase of $\psi$. 

We are now able to enlighten the meaning of the choice ${\cal S}_{0}=2 m {\cal D}$. We have seen that it is only under this particular choice that the fundamental equation of dynamics can be integrated. If we do not make this choice, the $\psi$ function is a solution of a third-order, non-linear equation such that no precise physical meaning can be given to it. We therefore claim that this choice has a profound physical significance, since the meaning of $\psi$ is directly related to the fact that it is a solution of the Schr\"odinger equation. In this equation, ${\cal S}_{0}$ is nothing but the fundamental action constant $\hbar$, while ${\cal D}$ defines the fractal/non-fractal transition (i.e. the transition from explicit scale dependence to scale independence), which is based on the constant $\lambda={2{\cal D}/c}$. Therefore, the relation ${\cal S}_{0}=2 m {\cal D}$ becomes a relation between mass and a length-scale, which can be written
\be
\lambda_{c}=\frac{\hbar}{mc} \; .
\label{eq.55}
\ee

We recognize here the definition of the Compton length. Its profound meaning - i.e. up to the fundamental constants $\hbar$ and $c$, that of inertial mass itself - is thus given, in our framework, by the transition scale from explicit scale dependence to scale independence. It will be completely enlightened in the relativistic-motion case. This length scale is to be understood as a structure of scale space, not of standard space. 

We recover, in this case, the standard form of Schr\"odinger's equation
\be
\frac{\hbar^2}{2m} \Delta \psi + i \hbar \frac{\partial}{\partial t}\psi = 
\Phi \psi .
\label{eq.56}
\ee

New insight about the statistical meaning of the wavefunction (Born postulate) can also be gained from the very construction of the theory. Indeed, while in standard quantum mechanics the existence of a complex wavefunction is set as a founding axiom, in the scale relativity framework it is related to the twin velocity field of the infinite family of geodesics with which the ``particle" is identified, through Eq.~(\ref{eq.41}), ${\cal V}=-2i{\cal D} \nabla(\ln \psi)$. This opens the possibility of getting a derivation of Born's postulate in this context. 

This question has already been considered by Hermann \cite{RH97}, who obtained numerical solutions of the equation of motion (\ref{eq.37}) in terms of a large number of explicit trajectories (in the case of a free particle in a box). He constructed a probability density from these trajectories and recovered in this way solutions of the Schr\"odinger equation without writing it and without using a wavefunction. 

Let us indeed derive the Born postulate in the one-dimensional stationary case. Consider the velocity field given by the real part $V$ of the complex velocity ${\cal V}$. It has been defined in such a way that it is identified with the classical velocity at the classical limit. Since this is a purely geometrical ``object", the density of the fluid corresponding to this velocity field is a probability density $P$ which satisfies the continuity equation, ${\partial P /  \partial t} + \mbox{div} (P V) = 0$. 

Let us now set $\rho = \psi \psi^{\dag}$, i.e. $\psi=\sqrt{\rho} \exp (iS/2m{\cal D})$. By replacing $\psi$ by this expression in the Schr\"odinger equation (\ref{eq.54}), and by using the identity $V=2 {\cal D} \nabla S$, its imaginary part now reads
\be
{\partial\rho\over \partial t} + \mbox{div} (\rho V) = 0.
\label{eq.57}
\ee
Since $P$ and $\rho$ are both solutions of a continuity equation, one can easily prove that their ratio $K=P/\rho$ is a solution of the equation $dK/dt=\partial K/ \partial t + V. \nabla K=0$. In the stationary one-dimensional case it is solved as $K = $constant, and therefore the squared modulus of the wavefunction is identified with the probability density after a proper normalization. The general time-dependent three-dimensional case will be considered in a forthcoming work.

\section{Klein-Gordon equation}
\label{s:ckgeq}

In Sec.~\ref{s:schro}, the Schr\"odinger equation has been derived as a geodesic equation in a fractal three-space, for non-relativistic motion, in the framework of Galilean scale relativity. In the rest of this paper we shall be concerned with relativistic motion in the same framework of Galilean scale relativity and shall derive the 
corresponding free-particle quantum mechanical equations (Klein-Gordon's and Dirac's) as geodesic equations in a four-dimensional fractal spacetime. 

\subsection{Relativistic-motion covariant derivative operator}
\label{ss:mrcdo}

Most elements of the approach summarized in Sec.~\ref{s:schro} remain 
correct in the relativistic-motion case, with the time, $t$, replaced by the proper time, $s$, as the curvilinear parameter along the geodesics. Now, not only space, but the full spacetime continuum is considered to be non-differentiable, thus fractal. We consider a small increment $dX_{\mu}$ of a non-differentiable four-coordinate along one of the geodesics of the fractal spacetime. We can, as above, decompose $dX^{\mu}$ in terms of a classical part $\overline{{\cal C} \ell}\langle dX^{\mu}\rangle =dx^{\mu}=v_{\mu}ds$ and a fluctuation respective to this classical part, $d\xi^{\mu}$, such that $\overline{{\cal C} \ell}\langle d\xi^{ \mu }\rangle=0$, by definition. The non-differentiable nature of spacetime implies the breaking of the reflection invariance at the infinitesimal level. If one reverses the sign of the proper time differential element, the classical part of the velocity  $v _{+}$ becomes $v _{- }$. We therefore consider again both the $(ds > 0)$ and $(ds < 0)$ processes on the same footing. Then the information needed to describe the system is doubled with respect to the standard differentiable description and can be once more accounted for by the use of complex numbers. The new complex process, as a whole, recovers again the fundamental property of microscopic reversibility. 

The elementary displacement along a geodesic of fractal dimension $D=2$, respectively, for the $(+)$ and  $(- )$ processes, is then
\begin{equation}
\label{1.}
dX _{\pm } ^{\mu } =  d _{\pm }x ^{\mu }  +  d\xi _{\pm } ^{\mu }  =  
 v _{\pm } ^{\mu } ds  + u _{\pm }^{\mu }\sqrt{2 {\cal D}} ds ^{1/2},
\end{equation}
with $d _{\pm }x ^{\mu }  = v _{\pm } ^{\mu } d s$, $d\xi_{\pm }^{ \mu}  =u _{\pm }^{\mu }\sqrt{2 {\cal D}} ds ^{1/2}$ and $u _{\pm } ^{\mu } $, a dimensionless fluctuation. The length-scale $2 {\cal D}$ is introduced for dimensional purpose. We define the classical derivatives, $d/ds _{+}$ and $d/ds_{-}$, using the classical part extraction procedure exposed in Sec.~\ref{ss:scderiv}, as
\be
\label{2.}
\frac{d}{ds _{\pm }} f(s)  =  \lim _{\delta s\rightarrow 0\pm} \; 
\overline{{\cal C} \ell}\left\langle\frac{f(s +\delta s ) - f(s)}{\delta s} 
\right\rangle .   
\ee

Once applied to $x^\mu $, they yield two classical four-velocities 
\begin{equation}
\label{3.}
  \frac{d}{ds _{+}} x ^{\mu  }(s)  = v ^{\mu } _{+}  \qquad 
\frac{d}{ds _{- }} x ^{\mu  }(s)  = v ^{\mu } _{-}. 
\end{equation}

They can be combined to construct a complex derivative operator
\begin{equation}
\frac{\dfr}{ds} = {1\over 2} \left( \frac{d}{ds_+} + \frac{d}{ds_-} \right) 
- {i\over 2} \left(\frac{d}{ds_+} - \frac{d}{ds_-}\right) .
\label{4.}
\end{equation}

The application of this operator to the position vector yields a complex four-velocity 
\begin{equation}
\label{5.}
{\cal V}  ^{\mu } =    \frac{\dfr}{ds}  x ^{\mu }   =   V ^{\mu  }- i  
U ^{\mu  } =    \frac{v ^{\mu } _{+ }+ v ^{\mu } _{- }}{2}   -  i    
\frac{v ^{\mu } _{+ }- v ^{\mu } _{- }}{2} \;  .   
\end{equation}

As regards the fluctuations, the generalization to four dimensions of the fractal behaviour described in Eq.~(\ref{1.}) gives
\begin{equation}
\label{6.}
\overline{{\cal C} \ell}\langle d\xi  ^{\mu } _{\pm }d\xi  ^{\nu } _{\pm }
\rangle =  \mp  2{\cal D} \eta  ^{\mu \nu } ds   .   
\end{equation}

As noted in Sec.~\ref{ss:scderiv}, each term on the left-hand side 
product behaves as an independent fractal fluctuation around its own 
classical part, which justifies a mathematical treatment analogous to that of a stochastic Wiener process. We make in the present paper the choice of a $(+,-,-,-)$ signature for the Minkowskian metric of the classical spacetime, $\eta  ^{\mu \nu }$. Now, a diffusion (Wiener) process makes sense only in $R ^{4}$ where the ``metric'' $\eta  ^{\mu \nu }$ should be positive definite, if one wants to interpret the continuity equation satisfied by the probability density as a 
Kolmogorov equation \cite{RH68}. Several proposals have been made to solve this problem \cite{DG85,MS88}, which are equivalent in the end, and amount to transforming a Laplacian operator in $R ^{4}$ into a Dalembertian. Namely, the two differentials of a function $f({x},{ s})$ may be written
\begin{equation}
\label{7.}
{df\over {ds _{\pm }}} =  \left ({\partial \over {\partial s}}  + v ^{\mu } 
_{\pm  } \partial  _{\mu }   \mp   {\cal D}   \partial   ^{\mu }
\partial  _{\mu } \right ) f    .   
\end{equation}

As we only consider $s$-stationary functions, not explicitly depending on the proper time $s$, the complex covariant derivative operator reduces to
\begin{equation}
\label{8.}
\frac {\dfr}{ds}   =    ({\cal V}  ^{\mu }  +  i  {\cal D}  
\partial   ^{\mu } ) \partial  _{\mu }   .        
\end{equation}

The plus sign in front of the Dalembertian comes from the choice of the metric signature.

\subsection{Geodesic equation}
\label{ss:csap}

To write the equation of motion, we use a generalized equivalence principle (identical to a strong covariance principle). We therefore obtain a geodesic equation in terms of the covariant derivative 
\be
\frac {\dfr {\cal V}_{\nu}}{ds}=0
\label{9.}
\ee

We now introduce the complex action according to
\be
d{\cal S}=\partial_{\nu}{\cal S}\; dx^{\nu}=-mc \; {\cal V}_{\nu} \; dx^{\nu}
\label{10.}
\ee

The complex four-momentum can thus be written as
\be
{\cal P}_{\nu}=mc \; {\cal V}_{\nu}= -\partial_{\nu}{\cal S}.
\label{eq.86}
\ee

Now, the complex action, ${\cal S}$, characterizes completely the dynamical state of the particle, and we can introduce a complex wavefunction
\begin{equation}
\label{12.}
\psi    =   e  ^{i{\cal S} /{\cal S}_0 }.    
\end{equation}

It is linked to the complex four-velocity by Eq.~(\ref{eq.86}), which gives
\be
{\cal V} _{\nu } = {{i {\cal S}_0\over {mc}}} \partial_{\nu }\ln\psi.
\label{13.}
\ee

\subsection{Free-particle Klein-Gordon equation}
\label{ss:ckg}

Replacing, in Eq.~(\ref{9.}), the covariant derivative by its expression given by Eq.~(\ref{8.}) and the complex four-velocity by that of Eq.~(\ref{13.}), we obtain 
\be
-{{\cal S}_0^2\over {m^2c^2}}\partial^{\mu}\ln\psi \partial_{\mu}\partial_{\nu}
\ln\psi - {{\cal S}_0 {\cal D}\over {mc}}\partial^{\mu}\partial_{\mu}
\partial_{\nu}\ln\psi=0.
\label{14.}
\ee

The particular choice, ${\cal S}_0=\hbar=2mc{\cal D}$, analogous 
to the one discussed in Sec.~\ref{ss:gnscheq}, allows us to simplify the left-hand side of Eq.~(\ref{14.}), using the following identity (which generalizes its three-dimensional counterpart of Eq.~(\ref{eq.50}))
\begin{equation}
\frac{1}{2}  \partial   ^{\nu } _{  }\left(  \frac{\partial _{\mu }\partial 
^{\mu } \psi }{\psi } \right) = \left( \partial _{\mu } \ln\psi + 
\frac{1}{2} \partial _{\mu } \right) \partial ^{\mu } \partial^{\nu }\ln\psi . 
\end{equation}

Dividing by the constant factor ${\cal D}^2$, we obtain the equation of motion of the free particle under the form
\begin{equation}
\label{14bis.}
 \partial ^{\nu } \left(  \frac{\partial ^{\mu }\partial 
_{\mu }\psi}{\psi} \right)     =  0   .    
\end{equation}

Therefore, the Klein-Gordon equation (without electromagnetic field)
\begin{equation}
\label{15.}
\partial ^{\mu } \partial _{\mu}\psi + {m ^{2 }c ^{2}\over {\hbar^2}}\psi=0,  
\end{equation}
becomes an integral of motion of the free particle, provided the integration constant is chosen equal to a squared mass term, $m^2c^2/\hbar^2$. The quantum behaviour described by this equation and the probabilistic interpretation given to $\psi $ is here reduced to the description of a free fall in a fractal spacetime, in analogy with Einstein's general relativity where a particle subjected to the effect of gravitation is described as being in free fall in a curved spacetime. 

Another important property is the way the characteristic length of the 
fractal (quantum) to differentiable (classical) transition (the Compton length of the particle in its rest frame) and the mass term in the Klein-Gordon equation appear. Here, as in the non-relativistic-motion case (Schr\"odinger), we identify the 
constant $2{\cal D}$ with the Compton length $\hbar/mc$, provided ${\cal S}_0$ is the fundamental action constant $\hbar$, in order to obtain the motion equation under the form of a vanishing four-gradient. As for the mass term in the final Klein-Gordon equation, it appears as a mere integration constant. However, it becomes connected to the transition length in the rest frame when matching these results with the non-relativistic description.

\section{Bi-quaternionic Klein-Gordon equation}
\label{s:kgeq}

It has been known for a long time that the Dirac equation proceeds from the Klein-Gordon equation when written in a quaternionic form \cite{CL29,AC37}. However, no physical reason was known for this important mathematical property. We propose in the current section to introduce naturally, as a consequence of the non-differentiable geometry, a bi-quaternionic covariant derivative operator, leading to the definition of a bi-quaternionic velocity and wavefunction, which we use to derive the Klein-Gordon equation in a bi-quaternionic form. We use the quaternionic formalism, as introduced by Hamilton \cite{WH66}, and further developed by Conway \cite{AC37,AC45} (see also Synge \cite{JS72} and Scheffers \cite{GS93}).

\subsection{Symmetry breakings and bi-quaternionic covariant derivative operator}
\label{ss:bqdop}

Most of the approach described in Sec.~\ref{s:ckgeq} remains applicable. However, the main new features obtained in the now studied case proceed from a deeper description of the scale formalism. Namely we now consider a more general case involving the subsequent breaking of the symmetries: \\
$ds\leftrightarrow -ds$ \\
$dx^{\mu}\leftrightarrow -dx^{\mu}$ \\
$x^{\mu}\leftrightarrow -x^{\mu}$. 

Indeed, we have up to now considered only the effect of non-differentiability on the total derivative $d/ds$ (that amounts to $d/dt$ in the non-relativistic case leading to the Schr\"odinger equation). Now the velocity fields of the geodesic bundles are functions of the coordinates, so that we are led to analyse also the physical meaning of the partial derivatives $\partial/\partial x$ (we use only one coordinate variable in order to simplify the writing) in the decomposition $d/ds=\partial/\partial s+(dx/ds) \partial/\partial x$. Strictly speaking, $\partial f / \partial x$ does not exist in the non-differentiable case. We are therefore once again led to introduce fractal functions $f(x,\delta x)$, explicitly dependent on the coordinate resolution interval, whose derivative is undefined only at the unobservable limit $\delta x \rightarrow 0$. As a consequence of the very construction of the derivative, which needs two points to be defined (instead of one for the position and time coordinates), there are two definitions of the partial derivative of a fractal function instead of one, namely
\be
{{\partial f }\over {\partial x}}_{+} = {{f(x+dx,dx)-f(x,dx)}\over {dx}},
\label{eq.57b}
\ee
\be
{{\partial f} \over {\partial x}}_{-} = {{f(x,dx)-f(x-dx,dx)}\over {dx}}.
\label{eq.57c}
\ee
They are transformed one into the other under the reflection $dx \leftrightarrow -dx$. 

The $x^{\mu}\leftrightarrow -x^{\mu}$ symmetry corresponds to the parity $P$ and time-reversal $T$ symmetries the breaking of which is already taken into account in the standard definition of Dirac spinors in terms of pairs of Pauli spinors \cite{LL72}. 

We have already stressed, in Sec.~\ref{ss:fracsp}, that differentiability can be recovered (at large scales), even when dealing with non-differentiability. However, this implies that, for any set of differential equations describing a given process, the physical and mathematical descriptions are only coincident on a limited scale range. We can say that any consistent mathematical tool 
lives in a description space which is tangent to the physical space and that the validity of this tool is therefore limited to a finite scale region around the contact point. For the Schr\"odinger and complex Klein-Gordon equations, the tangent mathematical space is such that $dt=\delta t$ (Schr\"odinger) or $ds=\delta s$ (Klein-Gordon). When jumping to smaller (higher energy) scales, we are led to give up this peculiar choice, and retain, at the Dirac scale, a new mathematical description where the differentials and the resolution variables do no longer coincide. However, this description is only 
valid at the scales where the Dirac equation applies without corrections and we shall be led to improve it when going to yet smaller scales. 

In the scaling domain, the four spacetime coordinates $X^\mu(s, \epsilon_{\mu}, \epsilon_{s})$ are fractal functions of the 
proper time, $s$, and of the resolutions, $\epsilon_\mu$ and $\epsilon_s$. We consider the case when, for an elementary displacement $dX^\mu$ corresponding to a shift $ds$ in the curvilinear parameter, the resolutions verify $\epsilon_{\mu}<dX^{\mu}$ and 
$\epsilon_{s}<ds$. 

We can apply, to these two different elementary displacements, the reasoning developed in Secs.~\ref{ss:difftsb} and ~\ref{ss:scderiv} which yields, considering first the displacement $dX^{\mu}$, the canonical decomposition
\be
dX^{\mu}=dx^{\mu}+d\xi^{\mu},
\label{eq.58}
\ee
with
\be 
\overline{{\cal C} \ell}\langle dX^\mu\rangle=dx^{\mu} = 
v^{\mu}_{+\atop{\mu}} \; ds,
\label{eq.59}
\ee
\be
d\xi^{\mu}=a^{\mu}_+ \sqrt{2 \cal{D}} (ds^{2})^{\frac{1}{2D}}, \qquad 
\overline{{\cal C} \ell}\langle a^{\mu}_+\rangle=0, \qquad \overline{{\cal C} 
\ell}\langle (a^{\mu}_+)^2 \rangle=1,
\label{eq.60}
\ee
and
\be
-dX^{\mu}=\delta x^{\mu}+\delta \xi^{\mu},
\label{eq.61}
\ee
with
\be 
\overline{{\cal C} \ell}\langle -dX^\mu\rangle=\delta x^{\mu} = 
v^{\mu}_{-\atop{\mu}} \; ds,
\label{eq.62}
\ee
\be
\delta \xi^{\mu}=a^{\mu}_- \sqrt{2 \cal{D}} (ds^{2})^{\frac{1}{2D}}, \qquad 
\overline{{\cal C} \ell}\langle a^{\mu}_-\rangle=0, \qquad \overline{{\cal C} 
\ell}\langle (a^{\mu}_-)^2 \rangle=1.
\label{eq.63}
\ee

In the differentiable case, $dX^\mu=-(-dX^\mu)$, and therefore 
$v^{\mu}_{+\atop{\mu}}=-v^{\mu}_{-\atop{\mu}}$. This is no longer the 
case in the non-differentiable case, where the local symmetry 
$dx^{\mu}\leftrightarrow -dx^{\mu}$ is broken. 

Furthermore, we must also consider the breaking of the symmetry 
$ds\leftrightarrow -ds$ proceeding from the twofold definition of the 
derivative with respect to the curvilinear parameter $s$. Applied to $X^\mu$, considering an elementary displacement $ds$, the classical part extraction process gives two classical derivatives $d/ds_+$ and $d/ds_-$, which yield in turn two classical velocities, which we denote by $v^\mu_{{\pm \atop{s}} {+\atop{\mu}}}$. Considering now the same extraction process, applied to an elementary displacement $-ds$, the classical derivatives $d/ds_+$ and $d/ds_-$ allow us again to define two classical velocities, denoted by $v^\mu_{{\pm \atop{s}} {-\atop{\mu}}}$. We summarize this result as 
\be
v^{\mu}_{{\pm \atop{s}} {+\atop{\mu}}}={dx^\mu\over {ds_{\pm}}}, \qquad 
v^{\mu}_{{\pm \atop{s}} {-\atop{\mu}}}={\delta x^\mu\over {ds_{\pm}}} \; .
\label{eq.64}
\ee

We can, at this stage, define several total derivatives with respect 
to $s$ of a fractal function $f$. We write them - using a compact 
straightforward notation with summation over repeated indices - after 
substituting, in the four-dimensional analogue of Eq.~(\ref{eq.30}), the expressions for the derivatives of $X^\mu$ which are obtained when using Eqs.~(\ref{eq.58}) to (\ref{eq.63}) with the expressions of Eq.~(\ref{eq.64}) for the classical velocities
\be
{df\over {ds}}{}_{{\pm \atop{s}} {\pm \atop{x}} {\pm \atop{y}} 
{\pm \atop{z}} {\pm \atop{t}}}= {\partial f\over {\partial s}} + 
(v^{\mu}_{{\pm \atop{s}} {\pm \atop{\mu}}} + w^{\mu}_{{\pm \atop{s}} 
{\pm \atop{\mu}}}){\partial f\over {\partial X^\mu}} + a^\mu_\pm a^\nu_\pm 
{\cal D}{\partial^2 f\over {dX^\mu dX^\nu}} \; ,
\label{eq.65}
\ee
with
\be
w^\mu=a^\mu {\sqrt {2 {\cal D}}} (ds^2)^{{1\over {2D}}-{1\over 2}}.
\label{eq.66}
\ee

Now, when we apply the classical part operator to Eq.~(\ref{eq.65}), using Eqs.~(\ref{eq.60}), (\ref{eq.63}) and (\ref{eq.66}), the $w^\mu$ disappear at the first order, but, at the second order, for the fractal dimension $D=2$, we obtain the analogue of Eq.~(\ref{6.})
\be
\overline{{\cal C} \ell}\langle w^{\mu}_{{\pm \atop{s}} {\pm \atop{\mu}}} 
w^{\nu}_{{\pm \atop{s}} {\pm \atop{\mu}}}\rangle=\mp 2 {\cal D} 
\eta^{\mu \nu} ds ,
\label{eq.67b}
\ee
The $\mp$ sign on the right-hand side is the inverse of the $s$-sign on the left-hand side. We can thus write the classical part of the total derivatives of Eq.~(\ref{eq.65}) as
\be
{df\over {ds}}{}_{{\pm \atop{s}} {\pm \atop{x}} {\pm \atop{y}} 
{\pm \atop{z}} {\pm \atop{t}}}= \left({\partial \over {\partial s}} + 
v^{\mu}_{{\pm \atop{s}} {\pm \atop{\mu}}} {\partial _\mu} 
\mp {\cal D}{\partial^\mu \partial_\mu}\right) f \; ,
\label{eq.67c}
\ee
where the $\mp$ sign on the right-hand side is still the inverse of the $s$-sign. 

We apply these derivatives to the position vector $X^\mu$ and obtain, as expected,
\be
{dX^{\mu}\over {ds}}{}_{{\pm \atop{s}} {\pm \atop{\mu}}} = v^{\mu}_{{\pm 
\atop{s}} {\pm \atop{\mu}}} .
\label{eq.67}
\ee

We consider now the four fractal functions $-X^\mu(s, \epsilon_\mu, 
\epsilon_s)$. At this description level, there is no reason for 
$(-X^{\mu})(s, \epsilon_\mu, \epsilon_s)$ to be everywhere equal to 
$-(X^{\mu})(s, \epsilon_\mu, \epsilon_s)$, owing to a local breaking of the P and T symmetries. We can therefore apply, to the two different elementary displacements of these functions, the canonical decomposition and the subsequent reasoning applied above to the $X^\mu$ function, and obtain the following doublings of the classical velocity:
\be
{{\tilde d}X^{\mu}\over {ds}}{}_{{\pm \atop{s}} {\pm \atop{\mu}}} = 
{\tilde v}^{\mu}_{{\pm \atop{s}} {\pm \atop{\mu}}}.
\label{eq.77}
\ee

We use notation analogous to that employed for the $X^\mu$ case, with the addition of a tilde. If we consider the simplest peculiar case when the breaking of the symmetry $dx^{\mu}\leftrightarrow -dx^{\mu}$ is isotropic as regards the four spacetime coordinates (i.e. the signs corresponding to the four $\mu$ indices are chosen equal), we are left with eight non-degenerate components - four $v^{\mu}_{{\pm \atop{s}} {\pm \atop{\mu}}}$ and four ${\tilde v}^{\mu}_{{\pm \atop{s}} {\pm \atop{\mu}}}$ - which can be used to define a bi-quaternionic four-velocity. We write
\begin{eqnarray}
{\cal V}^\mu={1\over 2}(v^\mu_{++} + {\tilde v}^\mu_{--})-{i\over 2}
(v^\mu_{++} - {\tilde v}^\mu_{--}) +\left[{1\over 2}(v^\mu_{+-} + 
v^\mu_{-+})-{i\over 2}(v^\mu_{+-} - {\tilde v}^\mu_{++})\right] e_1 
\nonumber \\
+ \left[{1\over 2}(v^\mu_{--} + {\tilde v}^\mu_{+-})-{i\over 2}(v^\mu_{--} - 
{\tilde v}^\mu_{-+})\right] e_2 + \left[{1\over 2}(v^\mu_{-+} + 
{\tilde v}^\mu_{++})-{i\over 2}({\tilde v}^\mu_{-+} + 
{\tilde v}^\mu_{+-})\right] e_3 .
\label{eq.78}
\end{eqnarray}

The freedom in the choice of the actual expression for ${\cal V}^\mu$ is constrained by the following requirements: at the limit when $\epsilon_\mu \rightarrow dX^\mu$ and $\epsilon_s \rightarrow ds$, every $e_i$-term in Eq.~(\ref{eq.78}) goes to zero, and, as ${\tilde v}^\mu_{--} = v^\mu_{-+}$ in this limit, one recovers the complex velocity of Eq.~(\ref{eq.28}), ${\cal V}^\mu=[v^\mu_{++}+v^\mu_{-+}-i(v^\mu_{++}-v^\mu_{-+})]/2$; at the classical limit, every term in this equation vanishes, save the real term, and the velocity becomes classical, i.e. real: ${\cal V}^\mu=v^\mu_{++}$. 

The bi-quaternionic velocity thus defined corresponds to a bi-quaternionic derivative operator $\ddfr /ds$, similarly defined, and yielding, when applied to the position vector $X^\mu$, the corresponding velocity. For instance, the derivative operator attached to the velocity in Eq.~(\ref{eq.78}) is written
\begin{eqnarray}
{\ddfr \over {ds}}={1\over 2}\left({d\over {ds}}{}_{++}+{{\tilde d}\over 
{ds}}{}_{--}\right) -{i\over 2}\left({d\over {ds}}{}_{++} - {{\tilde d}\over 
{ds}}{}_{--}\right) + \left[{1\over 2}\left({d\over {ds}}{}_{+-} + {d\over 
{ds}}{}_{-+}\right) - {i\over 2}\left({d\over {ds}}{}_{+-} - {{\tilde d}\over 
{ds}}{}_{++}\right)\right] e_1 \nonumber \\
+ \left[{1\over 2}\left({d\over {ds}}{}_{--} + {{\tilde d}\over 
{ds}}{}_{+-}\right) - {i\over 2}\left({d\over {ds}}{}_{--} - {{\tilde d}\over 
{ds}}{}_{-+}\right)\right] e_2 + \left[{1\over 2}\left({d\over {ds}}{}_{-+}+
{{\tilde d}\over {ds}}{}_{++}\right) -{i\over 2}\left({{\tilde d}\over 
{ds}}{}_{-+} + {{\tilde d}\over {ds}}{}_{+-}\right)\right] e_3 . \nonumber \\
\label{eq.79}
\end{eqnarray}
Substituting Eq.~(\ref{eq.67c}) and its tilde counterpart into 
Eq.~(\ref{eq.79}), we obtain the expression for the bi-quaternionic 
proper-time derivative operator 
\be
{\ddfr \over {ds}}= [1+e_1+e_2+(1-i)e_3]{\partial \over {\partial s}} + 
{\cal V}^\mu \partial_\mu + i{\cal D} \partial ^\mu \partial _\mu ,
\label{eq.80}
\ee
the $+$ sign in front of the Dalembertian proceeding from the choice of the metric signature $(+,-,-,-)$. We keep here, for generality, the $\partial / \partial s$ term, stressing that it is actually of no use, since the various physical functions are not explicitly dependent on $s$. Therefore, the form that we finally obtain for the bi-quaternionic derivative operator is unchanged with respect to the previous complex operator: this is another manifestation of the covariance of the scale relativity description. It is easy to check that this operator, applied to the position vector $X^\mu$, gives back the bi-quaternionic velocity ${\cal V}^\mu$ of Eq.~(\ref{eq.78}). 

The expression that we have written for ${\cal V}^\mu$ in Eq.~(\ref{eq.78}) is one among the various choices we could have retained to define the bi-quaternionic velocity. The main constraint limiting this choice is the recovery of the complex and real velocities at the non-relativistic-motion and classical limits. We also choose ${\cal V}^{\mu}$ such as to obtain the third term on the right-hand side of Eq.~(\ref{eq.80}) under the form of a purely imaginary Dalembertian, which allows us to recover an integrable equation of motion. To any bi-quaternionic velocity satisfying both prescriptions there corresponds a bi-quaternionic derivative operator  $\ddfr /ds$, similarly defined, and yielding this velocity when applied to the position vector $X^\mu$. But, whatever the definition retained, the derivative operator keeps the same form in terms of the bi-quaternionic velocity ${\cal V}^\mu$, as given by Eq.~(\ref{eq.89}). Therefore, the different choices allowed for its definition merely correspond to different mathematical representations leading to the same physical result.

\subsection{Bi-quaternionic geodesic equation}
\label{ss:bqsap}

We now apply the generalized equivalence principle, as developed in 
Sec. \ref{ss:csap}. The free-motion equation issued from this principle is the geodesic equation 
\be
{\ddfr {\cal V}_{\mu} \over {ds}}=0,
\label{eq.84b}
\ee
where ${\cal V}_{\mu}$ is the bi-quaternionic four-velocity, e.g., as defined in Eq.~(\ref{eq.78}). 

We introduce the bi-quaternionic action according to
\be
\delta {\cal S}= \partial_{\mu}{\cal S} \; dx^{\mu}=- mc \; {\cal V}_{\mu} \; 
\delta x^{\mu}.
\label{eq.85b}
\ee

We thus obtain the bi-quaternionic four-momentum, as
\be
{\cal P}_{\mu}=mc \; {\cal V}_{\mu}= -\partial_{\mu}{\cal S}.
\label{eq.86b}
\ee

We can now define a bi-quaternionic wavefunction, which is again a re-expression of the action and which we write
\be
\psi^{-1} \partial_{\mu} \psi = {i\over {c S_0}} \partial_{\mu} {\cal S},
\label{eq.87b}
\ee
using, on the left-hand side, the quaternionic product. This gives for the bi-quaternionic four-velocity, as derived from Eq.~(\ref{eq.86b}),
\be
{\cal V}_{\mu}=i{S_0\over m} \psi^{-1} \partial_{\mu} \psi.
\label{eq.88b}
\ee 

We could choose, for the definition of the wavefunction in Eq.~(\ref{eq.87b}), a commutated expression on the left-hand side, i.e. $(\partial_{\mu} \psi) \psi^{-1}$ instead of $\psi^{-1} 
\partial_{\mu} \psi$. But with this reversed choice, owing to the non-commutativity of the quaternionic product, we could not obtain the motion equation as a vanishing four-gradient, as in Eq.~(\ref{eq.95}). Therefore, we retain the above choice as the simplest one, i.e. yielding an equation which can be integrated.

\subsection{Free-particle bi-quaternionic Klein-Gordon equation}
\label{ss:fpkg}

We can now replace the covariant derivative operator $\ddfr /ds$ by its expression in Eq.~(\ref{eq.80}). As we only consider $s$-stationary functions, i.e. functions which do not explicitly depend on the proper time $s$, this operator reduces to
\be
{\ddfr \over {ds}}={\cal V}^\nu \partial_\nu + i{\cal D} \partial ^\nu 
\partial _\nu.
\label{eq.89}
\ee

The equation of motion (\ref{eq.84b}) is thus written 
\be
\left ({\cal V}^\nu \partial_\nu + i{\cal D} \partial ^\nu 
\partial _\nu \right ) {\cal V}_{\mu} = 0.
\label{eq.90}
\ee

Replacing ${\cal V}_{\mu}$, respectively ${\cal V}^{\nu}$, by their 
expressions in Eq.~(\ref{eq.88b}), we obtain 
\be
i{S_0\over m} \left ( i{S_0\over m} \psi^{-1} \partial^{\nu} \psi 
\partial_\nu + i{\cal D} \partial ^\nu \partial _\nu \right ) \left ( 
\psi^{-1} \partial_{\mu} \psi \right ) = 0.
\label{eq.91}
\ee

As in Sec. \ref{ss:gnscheq}, the choice ${\cal S}_0= 2 m {\cal D}$ allows us to simplify this equation as
\be
\psi^{-1} \partial^{\nu} \psi \; \partial_\nu (\psi^{-1}  
\partial_{\mu} \psi) + {1\over 2} \partial ^\nu \partial _\nu (\psi^{-1} 
\partial_{\mu} \psi) = 0.
\label{eq.92}
\ee

The definition of the inverse of a quaternion
\be
\psi \psi^{-1} =  \psi^{-1} \psi = 1,
\label{eq.93}
\ee
implies that $\psi$ and $\psi^{-1}$ commute. But this is not necessarily the case for $\psi$ and $\partial _{\mu} \psi^{-1}$ nor for $\psi^{-1}$ and $\partial _{\mu} \psi$ and their contravariant counterparts. However, when we differentiate Eq.~(\ref{eq.93}) with respect to the coordinates, we obtain
\begin{eqnarray}
\psi \; \partial _{\mu} \psi^{-1} &=& - (\partial _{\mu} \psi) \psi^{-1} 
\nonumber \\
\psi^{-1} \partial _{\mu} \psi &=& - (\partial _{\mu} \psi^{-1}) \psi,
\label{eq.94}
\end{eqnarray}
and identical formulae for the contravariant analogues. 

Developing Eq.~(\ref{eq.92}), using Eqs.~(\ref{eq.94}) and the property $\partial^{\nu}\partial_{\nu} \partial_{\mu} = \partial_{\mu}\partial^{\nu} \partial_{\nu}$, we obtain, after some calculations,
\be
\partial_{\mu}[(\partial^{\nu}\partial_{\nu} \psi) \psi^{-1}] = 0.
\label{eq.95}
\ee

We integrate this four-gradient and write
\be
(\partial^{\nu}\partial_{\nu} \psi) \psi^{-1} + C = 0 ,
\label{eq.96}
\ee
of which we take the right product by $\psi$ to obtain 
\be
\partial^{\nu}\partial_{\nu} \psi + C \psi = 0.
\label{eq.97}
\ee

We therefore recognize the Klein-Gordon equation for a free particle with a mass $m$ (so that $m^2c^2/{\hbar}^2=C$), but now generalized to complex quaternions.

\section{Dirac equation}
\label{s:dieq}

We here use a long-known property of the quaternionic formalism, which allows to obtain the Dirac equation for a free particle as a mere square root of the Klein-Gordon operator (see, e.g., \cite{CL29,AC37}). We first develop the Klein-Gordon equation, as
\be
{1\over {c^2}}{\partial^2 \psi \over {\partial t^2}} = {\partial^2 \psi \over 
{\partial x^2}} + {\partial^2 \psi \over {\partial y^2}} + 
{\partial^2 \psi \over {\partial z^2}} - {m^2c^2\over {\hbar^2}} \psi.
\label{eq.98}
\ee

Thanks to the property of the quaternionic and complex imaginary units 
$e^2_1=e^2_2=e^2_3=i^2=-1$, we can write Eq.~(\ref{eq.98}) under the form
\be
{1\over {c^2}}{\partial^2 \psi \over {\partial t^2}} = e^2_3{\partial^2 \psi 
\over {\partial x^2}}e^2_2 + ie^2_1{\partial^2 \psi \over {\partial y^2}}i + 
e^2_3{\partial^2 \psi \over {\partial z^2}}e^2_1 + i^2{m^2c^2\over {\hbar^2}} 
e^2_3 \psi e^2_3.
\label{eq.99}
\ee

We see that Eq.~(\ref{eq.99}) is obtained by applying twice to the bi-quaternionic wavefunction $\psi$ the operator $\partial/c\partial t$ written as
\be
{1\over c}{\partial\over {\partial t}} = e_3 {\partial \over 
{\partial x}}e_2 + e_1 {\partial \over 
{\partial y}}i + e_3 {\partial \over {\partial z}}e_1 - 
i{mc\over \hbar}e_3 ( \quad ) e_3.
\label{eq.101}
\ee

The three Conway matrices $e_3(\quad)e_2$, $e_1(\quad)i$ and 
$e_3(\quad)e_1$ \cite{AC45}, figuring in the right-hand side of Eq.~(\ref{eq.101}), can be written in the compact form $-\alpha^k$, 
with 
\begin{displaymath}
\alpha^k= \left (
          \begin{array}{cc}
           0 & \sigma_k \\ 
           \sigma_k & 0 
           \end{array} 
           \right ), 
\end{displaymath}
the $\sigma_k$ being the three Pauli matrices, while the Conway matrix
\begin{displaymath}
e_3(\quad)e_3= \left (
          \begin{array}{cccc}
           1 & \quad 0 & \; \; 0 & 0 \\ 
           0 & \quad 1 & \; \; 0 & 0 \\
           0 & \quad 0 & \; \; -1 & 0 \\
           0 & \quad 0 & \; \; 0 & -1 
           \end{array}
           \right ) 
\end{displaymath}
is recognized as the Dirac $\beta$ matrix. We can therefore write 
Eq.~(\ref{eq.101}) as the non-covariant Dirac equation for a free fermion
\be
{1\over c}{\partial \psi \over {\partial t}} = - \alpha^k{\partial \psi \over 
{\partial x^k}} - i{mc\over \hbar}\beta \psi .
\label{eq.102}
\ee

The covariant form, in the Dirac representation, can be recovered by 
applying $ie_3(\quad)e_3$ to Eq.~(\ref{eq.102}). 

The isomorphism which can be established between the quaternionic and  
spinorial algebrae through the multiplication rules applying to the Pauli spin matrices allows us to identify the wavefunction $\psi$ 
with a Dirac spinor. Spinors and quaternions are both a representation of the SL(2,C) group ( see Ref. \cite{PR64} for a detailed discussion of the spinorial properties of bi-quaternions).

\section{Pauli equation}
\label{s:paeq}

Finally it is easy to derive the Pauli equation, since it is known that it can be obtained as a non-relativistic-motion approximation of the Dirac equation \cite{LL72}. Two of the components of the Dirac bi-spinor become negligible when $v<<c$, so that they become Pauli spinors (i.e. in our representation the bi-quaternions are reduced to quaternions) and the Dirac equation is transformed into a Schr\"odinger equation for these spinors with a magnetic dipole additional term. Such an equation is just the Pauli equation. Therefore, the Pauli equation is understood in the scale-relativistic framework as a manifestation of the fractality of space (but not time), while the symmetry breaking of space differential elements is nevertheless at work. 

\section{Conclusion}

Four fundamental motion equations of standard microphysics have been recovered, in the framework of Galilean scale relativity, as geodesic equations in a fractal space (Schr\"odinger and Pauli), then in a fractal spacetime (free Klein-Gordon and Dirac). It is interesting to note how the change from classical to quantum non-relativistic motion, then from quantum non-relativistic to quantum relativistic motion arises from successive symmetry breakings in the fractal geodesic picture. 

First, the complex nature of the wavefunction is the result of the 
differential (proper) time symmetry breaking, which is the simplest effect arising from the fractal structure of space (spacetime). It allows a statistical interpretation of this wavefunction in the form of the Born postulate (which is demonstrated in this framework in the stationary one-dimensional case). At this stage, Galilean scale relativity with a complex wavefunction permits the recovery of both the Schr\"odinger (Sec. \ref{s:schro}) and Klein-Gordon (Sec. \ref{s:ckgeq}) equations. 

To go on with the description of the elementary properties encountered 
in the microphysical world, we consider further breakings of 
fundamental symmetries, namely the differential coordinate symmetry 
($dx^{\mu} \leftrightarrow -dx^{\mu}$) breaking and the parity and 
time reversal symmetry breakings. They provide a four-complex-component wavefunction (i.e. a eight component wavefunction), of which the most natural mathematical representation is in terms of bi-quaternionic numbers. We therefore obtain the spinorial and the particle-anti-particle nature of elementary objects which we can describe as Dirac spinors. The Klein-Gordon equation can be recovered from a generalized equivalence principle in a bi-quaternionic form which naturally yields the free Dirac equation (Secs. \ref{s:kgeq} and \ref{s:dieq}). 

The Pauli equation, which is a non-relativistic motion approximation of the Dirac equation, then follows. It is understood, in the 
scale-relativistic framework, as a manifestation of the fractality of 
space (not time), while the symmetry breaking of the space differential elements is nevertheless at work (Sec. \ref{s:paeq}). 

It is worth stressing here that a non-differentiable and fractal spacetime is essentially non-local, since the ``particles" are identified with bundles of geodesics. Therefore, we recover the non-locality of the wavefunction of standard quantum mechanics. Moreover, having now derived the Dirac equation and the bi-spinor nature of the wavefunction exactly in its standard quantum mechanical form (there is no missing or additionnal variable in the non-differentiable spacetime representation, but only a change of variables with the same number of degrees of freedom), some profound aspects of quantum mechanics such as the EPR paradox and the breaking of Bell inequalities are also recovered in the new framework. \\

{\it Aknowledgments.} The authors thank Dr. T. Lehner and Pr. J.P. Connerade for their encouraging support and E. Seri\'e and the referees for useful remarks.

\appendix
\section{Complex numbers and bi-quaternions as a covariant representation}
\label{a:cqn}

\subsection{Introduction}

The first axioms of quantum mechanics state that one defines a state function (or probability amplitude) $\psi$ which is complex, is calculated as $\psi=\psi_{1}+\psi_{2}$ for two alternative channels, and is such that  the probability density is given by $\psi \psi^{\dagger}$ \cite{RF65}. Is it possible to derive them from more fundamental principles? Such an understanding, impossible in the framework of quantum mechanics itself (since these statements are its basic axioms), must be looked for in an enlarged paradigm. In scale relativity, we extend the founding stones of physics by giving up the hypothesis of spacetime differentiability. One of the main and simplest consequences is that the velocity field becomes two-valued, implying a two-valuedness of the Lagrange function and therefore of the action. Finally, the wavefunction is defined as a re-expression of the action, so that it will also be two-valued in the simplest case (see main text). 

But we have, up to now, admitted without justification that this two-valuedness is to be described in terms of complex numbers. Indeed, our equations are not simply a ``pasting" of real and imaginary equations, but involve the complex product from the very beginning of the calculations. In particular, the geodesic equation for a free particle takes the form of the equation of inertial motion
\begin{equation}
\label{appA}
\frac{\dfr^2}{dt^2} x^{k}=0,
\end{equation}
in terms of the ``covariant" complex time derivative  operator
\begin{equation}
\frac{\dfr}{dt} =\frac{\partial}{\partial t} + {\cal V} . \nabla - i {\cal D} \Delta.
\end{equation}

Equation (\ref{appA}) amounts to Schr\"odinger's equation (see main text). Since it corresponds to a second derivative, the complex product has mixed the real and imaginary quantities in a very specific way, so that it is not at all trivial that Schr\"odinger's equation is recovered at the end. 

The aim of this appendix is to address the question: what is the justification of using complex numbers, then bi-quaternions for representing the successive doublings of variables? As we shall see, complex numbers achieve a particular representation of quantum mechanics in terms of which the fundamental equations take their simplest form. Other choices for the representation of the two-valuedness and for the new product are possible, but these choices would give to the equations a more complicated form, involving additional terms (although the physical meaning would be unchanged). Bi-quaternions follow as a further splitting, at another level, of the complex numbers.

\subsection{Origin of complex numbers and quaternions in quantum mechanics}
\label{ss:cnq}

Let us return to the step of our demonstration where complex numbers are introduced. All we know is that each component of the velocity now takes two values instead of one. This means that each component of the velocity becomes a vector in a two-dimensional space, or, in other words, that the velocity becomes a two-index tensor. So let us introduce generalized velocities
\begin{equation}
\label{}
V^{k}_{\sigma}=(V^{k},U^{k})  \;\; ; \;\;  k= 1,2,3   \;\; ; \;\;  \sigma  =-,+ .
\end{equation}

This can be generalized to other physical quantities affected by the two-valuedness: scalars $A$ of the position space become vectors $A_{\sigma}$ of the new 2D-space, etc. While the generalization of the sum of these quantities is straighforward, $C^{k}_{\sigma} = A^{k}_{\sigma}+B^{k}_{\sigma}$, the generalization of the product is an open question at this stage. The problem amounts to finding a generalization of the standard product that keeps its most fundamental physical properties (e.g., internal composition law), when some of them cannot escape to be lost (e.g., commutativity when jumping to quaternions). 

From the mathematical point of view, we are here confronted with the well-known problem of the doubling of algebra (see, e.g.,  \cite{MP82}). The effect of the symmetry breaking $dt \leftrightarrow -dt$ (or $ds \leftrightarrow -ds$) is to replace the algebra ${\cal A}$, in which the classical physical quantities are defined, by a direct sum of two exemplars of ${\cal A}$, i.e. the space of the pairs $(a,b)$ where $a$ and $b$ belong to ${\cal A}$. The new vectorial space ${\cal A}^2$ must be supplied with a product in order to become itself an algebra (of doubled dimension). The same problem shows again when one takes also into account the symmetry breakings $dx^{\mu} \leftrightarrow -dx^{\mu}$ and $x^{\mu} \leftrightarrow -x^{\mu}$: this leads to new algebra doublings. The mathematical solution to this problem is well known: the standard algebra doubling amounts to supply ${\cal A}^2$ with the complex product. Then the doubling ${\rm I\! R^2}$ of ${\rm I\! R}$ is the algebra ${\rm I\! \!\!C}$ of complex numbers, the doubling ${\rm I\! \!\!C^2}$ of ${\rm I\!\! \!C}$ is the algebra ${\rm I\! H}$ of quaternions and the doubling ${\rm I\! H^2}$ of quaternions is the algebra of Graves-Cayley octonions. The problem is that the iterative doubling leads to a progressive deterioration of the algebraic properties. Namely the quaternion algebra is non-commutative, while the octonion algebra is also non-associative. But an important positive result for physical applications is that the doubling of a metric algebra is a metric algebra \cite{MP82}. 

These mathematical theorems lead us to use complex numbers, then quaternions, in order to describe the successive doublings due to discrete symmetry breakings at the infinitesimal level, which are themselves more and more profound consequences of spacetime non-differentiability. However, we give in what follows complementary arguments of a physical nature, which show that the use of the complex product in the first algebra doubling has a simplifying and covariant effect. 

In order to simplify the argument, let us consider the generalization of scalar quantities, for which the product law is the standard product in ${\rm I \!R}$. 

The first constraint is that the new product must remain an internal composition law. We also make the simplifying assumption that it remains linear in terms of each of the components of the two quantities to be multiplied. Therefore, we consider a general form for a bilinear internal product
\begin{equation}
\label{}
C^{\gamma}= A^{\alpha} \; \Omega_{\alpha \beta}^{\gamma}\; B^{\beta}
\end{equation}
where the matrix $\Omega_{\alpha \beta}^{\gamma}$ is a tensor (similar to the structure constants of a Lie group) that defines completely the new product. 

The second physical constraint is that we recover the classical variables and product at the classical limit. The mathematical equivalent is the requirement that ${\cal A}$ still be a sub-algebra of ${\cal A}^2$. Therefore, we identify $a_0 \in {\cal A}$ with $(a_0,0)$ and we set $(0,1)=\alpha$. This allows us to write the new two-dimensional vectors in the simplified form $a=a_{0}+a_{1} \alpha$, so that the product is now written
\begin{equation}
\label{}
c=( a_{0}+a_{1} \alpha)\,( b_{0}+b_{1} \alpha)= a_{0}b_{0}+a_{1}b_{1} \alpha^{2}+(a_{0}b_{1}+a_{1}b_{0})\alpha .
\end{equation}

The problem is now reduced to finding $\alpha^2$, i.e. only two $\Omega$ coefficients instead of eight.
\begin{equation}
\label{}
\alpha^{2}=\omega_{0}+\omega_{1} \alpha.
\end{equation}

Let us now return to the beginning of our construction. We have introduced two elementary displacements, each of them consisting of two terms, a classical part and a fluctuation (see Eq.~(\ref{eq.24}))
\begin{eqnarray} 
  dX_+(t) = v_+\;dt + d\xi_+(t) , \nonumber \\
  dX_-(t) = v_-\;dt + d\xi_-(t) .
\end{eqnarray}

Therefore, one can define velocity fluctuations $w_{+}=d\xi_{+}/dt$ and $w_{-}=d\xi_{-}/dt$, then a complete velocity in the doubled algebra \cite{LN99}
\begin{equation}
\label{}
{\cal V} + {\cal W} = \left(\frac{v_{+}+v_{-}}{2} - \alpha \, \frac{v_{+}-v_{-}}{2}\right) + \left(\frac{w_{+}+w_{-}}{2}-\alpha\, \frac{w_{+}-w_{-}}{2}\right) \; .
\end{equation}

In terms of standard methods, this writing would be forbidden since the velocity ${\cal W}$ is infinite from the viewpoint of usual differential calculus (it is $\propto dt^{-1/2}$). But we give meaning to this concept by considering it as an explicit function of the differential element $dt$, which becomes itself a variable. 

Now, from the covariance principle, the Lagrange function in the Newtonian case should strictly be written:
\begin{equation}
\label{}
{\cal L}= \frac{1}{2} m \; \overline{{\cal C} \ell}\langle({\cal V} + {\cal W})^{2}\rangle=\frac{1}{2} m \; \left(\overline{{\cal C} \ell}\langle{\cal V}^{2}\rangle + \overline{{\cal C} \ell}\langle{\cal W}^{2}\rangle\right)
\end{equation}

We have $\overline{{\cal C} \ell}\langle{\cal W}\rangle=0$, by definition, and $\overline{{\cal C} \ell}\langle{\cal V}{\cal W}\rangle=0$, because they are mutually independent. But what about $\overline{{\cal C} \ell}\langle{\cal W}^{2}\rangle$ ? The presence of this term would greatly complicate all the subsequent developments toward the Schr\"odinger equation, since it would imply a fundamental divergence of non-relativistic quantum mechanics. Let us expand it
\begin{eqnarray}
\label{}
4\overline{{\cal C} \ell}\langle{\cal W}^{2}\rangle&=&\overline{{\cal C} \ell}\langle  [(w_{+}+w_{-})-\alpha\, (w_{+}-w_{-})]^2 \rangle \nonumber \\
&=&\overline{{\cal C} \ell}\langle(w_{+}^2+w_{-}^2)(1+\alpha^2)-2\alpha(w_{+}^2-w_{-}^2) \nonumber \\
&+& 2 w_{+}w_{-}(1-\alpha^2)\rangle.
\end{eqnarray}

Since $\overline{{\cal C} \ell}\langle w_{+}^2\rangle=\overline{{\cal C} \ell}\langle w_{-}^2\rangle$ and $\overline{{\cal C} \ell}\langle w_{+}w_{-}\rangle=0$ (they are mutually independent), we finally find that $\overline{{\cal C} \ell}\langle{\cal W}^{2}\rangle$ can only vanish provided
\begin{equation}
\label{}
\alpha^2=-1,
\end{equation}
namely $\alpha=\pm i$, the imaginary. Therefore, we see that the choice of the complex product in the algebra doubling plays an essential physical role, since it allows the suppression of additional infinite terms in the final equations of motion.

\subsection{Origin of bi-quaternions}

A last point to be justified is the use of complex quaternions
(bi-quaternions) for describing the new algebra doublings that lead us to bi-spinors and the Dirac equation. One could think that the argument
given in Sec. \ref{ss:cnq} implies the use of Graves-Cayley octonions (and therefore the giving up of associativity) in the case of three successive doublings as considered in this paper. However,these three doublings are not on the same footing from a physical pointof view:

(i) The first two-valuedness comes from a discrete symmetry breaking at
the level of the differential invariant, namely $dt$ in the case of a
fractal space (yielding the Schr\"odinger equation) and $ds$ in the case of a fractal spacetime (yielding the Klein-Gordon equation). This
means that it has an effect on the total derivatives $d/dt$ and $d/ds$.
This two-valuedness is achieved by the introduction of complex 
variables.

(ii) The second two-valuedness (differential parity and time reversal 
violation, ``dX''), which is specifically introduced and studied in the present paper, comes from a new discrete symmetry breaking (expected from the giving up of the differentiability hypothesis) on the spacetime differential element $dx^{\mu} \leftrightarrow -dx^{\mu}$. It is subsequent to the first two-valuedness, since it has an effect on the partial derivative $\partial_{\mu}=\partial / \partial x^{\mu}$ that intervenes in the complex covariant derivative operator, namely,
\begin{equation}
\frac{\dfr}{ds}= ({\cal V}^{\mu}+i {\cal D} \partial^{\mu})\partial_{\mu}.
\end{equation}
(iii) The third two-valuedness is a standard effect of parity (P) and time reversal (T) in the relativistic-motion situation, which is not specific to our approach and is already used in the standard construction of Dirac spinors \cite{LL72}. It does not lead to a real information doubling, since Dirac spinors still have only four degrees of freedom as Pauli spinors do. Therefore, from the second and third doublings, complex numbers, then quaternions, can be introduced, which will affect variables which are already complex due to the first, more fundamental, doubling. This leads to the bi-quaternionic tool we use here. 

It is worth noting that these symmetry breakings are effective only at the level of the underlying description of elementary displacements (namely in the non-differentiable fractal spacetime). The effect of introducing a two-valuedness of variables in terms of double symmetrical processes amounts to recover symmetry in terms of the bi-process, and therefore in terms of the quantum tools which are built from it. In reverse, it opens a possible way of future investigation into the origin of other features characteristic of the microphysical world.

\subsection{Complex representation and covariant Euler-Lagrange equations.}

Let us confirm by another argument that our choice of a representation in terms of complex numbers for the two-valuedness of the velocity (namely in the minimal situation that leads to the Schr\"odinger equation) is a simplifying and covariant choice. We demonstrate hereafter that the standard form of the Euler-Lagrange equations is conserved when we combine the two velocities in terms of a unique complex velocity. 

In a general way, the Lagrange function is expected to be a function of the variables $x$ and their time derivatives $\dot{x}$. We have found that the number of velocity components $\dot{x}$ is doubled, so that we could have written
\begin{equation}
\label{app1}
L=L(x,\dot{x}_{+}, \dot{x}_{-},t).
\end{equation}

Instead, we have made the choice to write the Lagrange function as
\begin{equation}
\label{app2}
L=L(x,{\cal V},t).
\end{equation}

We now justify this choice by the covariance principle.
Re-expressed in terms of $\dot{x}_{+}$ and $\dot{x}_{-}$, the Lagrange 
function is written
\begin{equation}
L=L\left(x,\frac{1-i}{2} \; \dot{x}_{+} + \frac{1+i}{2} \; \dot{x}_{-},t\right).
\end{equation}

Therefore we obtain
\begin{equation}
\frac{\partial L}{\partial \dot{x}_{+}}=\frac{1-i}{2} \; \frac{\partial L}{\partial {\cal V}} \;\;\; ; \;\;\; \frac{\partial L}{\partial \dot{x}_{-}}=\frac{1+i}{2} \; \frac{\partial L}{\partial {\cal V}},
\end{equation}
while the new covariant time derivative operator reads
\begin{equation}
\frac{\dfr}{dt}=\frac{1-i}{2} \; \frac{d}{dt_{+}}+\frac{1+i}{2} \; \frac{d}{dt_{-}} \; .
\end{equation}

Let us write the stationary action principle in terms of the Lagrange 
function of Eq. (\ref{app1})
\begin{equation}
\delta S= \delta \int_{t_1}^{t_2} L(x,\dot{x}_{+}, \dot{x}_{-},t) \; dt = 0.
\end{equation}

It becomes
\begin{equation}
\int_{t_1}^{t_2}\, \left(  \frac{\partial L}{\partial x} \; \delta x +\frac{\partial L}{\partial \dot{x}_{+}} \; \delta  \dot{x}_{+} + \frac{\partial L}{\partial \dot{x}_{-}} \; \delta  \dot{x}_{-}  \right) \, dt = 0.
\end{equation}

Since $\delta  \dot{x}_{+} =d(\delta x)/dt_{+}$ and $\delta  \dot{x}_{-} =d(\delta x)/dt_{-}$, it takes the form
\begin{equation}
 \int_{t_1}^{t_2}\, \left(  \frac{\partial L}{\partial x} \; \delta x +\frac{\partial L}{\partial {\cal V} }\; \left[ \, \frac{1-i}{2} \; \frac{d}{dt_{+}} + \frac{1+i}{2} \; \frac{d}{dt_{-}}\,\right] \,  \delta  x  \right) \, dt = 0,
\end{equation}
i.e.
\begin{equation}
\int_{t_1}^{t_2}\, \left(      \frac{\partial L}{\partial x} \; \delta x      +     \frac{\partial L}{\partial {\cal V}} \; \frac{\dfr}{dt} \, \delta  x  \right)\, dt = 0.
\end{equation}

The subsequent demonstration of the Lagrange equations from the stationary action principle relies on an integration by parts. This integration by parts cannot be performed in the usual way without a specific analysis, because it involves the new covariant derivative. 

The first point to consider is the Leibniz rule for the covariant derivative operator $\dfr/dt$. Since $\dfr/dt=\partial/dt + {\cal V}\,.\, \nabla - i {\cal D} \Delta $ is a linear combination of first- and second-order derivatives, the same is true of its Leibniz rule. This implies an additional term in the expression for the derivative of a product \cite{JCP99A}:
\begin{equation}
\frac{\dfr}{dt} \left( \frac{\partial L}{\partial {\cal V}} \,.\, \delta x \right) =
\frac{\dfr}{dt}  \frac{\partial L}{\partial {\cal V}} \,.\, \delta x +  \frac{\partial L}{\partial {\cal V}}\,.\, \frac{\dfr}{dt} \delta x - 2 i \,{\cal D}\, \nabla\frac{\partial L}{\partial {\cal V}}\,.\, \nabla \delta x.
\end{equation}

Since $\delta x(t)$ is not a function of $x$, the additional term vanishes. Therefore, the above integral becomes
\be
\int_{t_1}^{t_2}\, \left[\left(      \frac{\partial L}{\partial x}    -   \frac{\dfr}{dt} \, \frac{\partial L}{\partial {\cal V}}   \right) \, \delta  x \, + \frac{\dfr}{dt} \left( \frac{\partial L}{\partial {\cal V}} \,.\, \delta x \right)\right] dt = 0.
\ee

The second point is the integration of the covariant derivative. We define a new integral as being the inverse operation of the covariant derivation, i.e. 
\be
\intfr \; \dfr f=f
\ee
in terms of which one obtains
\be
\intfr _{t_1}^{t_2} \; \dfr \left( \frac{\partial L}{\partial {\cal V}} \,.\, \delta x \right)= \left[ \frac{\partial L}{\partial {\cal V}} \,.\, \delta x \right]_{t_1}^{t_2}=0,
\ee
since $\delta x(t_1)=\delta x(t_2)=0$ by definition of the variation principle. Therefore, the action integral becomes
\begin{equation}
\int_{t_1}^{t_2}\, \left(      \frac{\partial L}{\partial x}    -   \frac{\dfr}{dt} \, \frac{\partial L}{\partial {\cal V}}   \right) \, \delta  x \, dt= 0.
\end{equation}

And finally we obtain generalized Euler-Lagrange equations that read
\begin{equation}
\frac{\dfr}{dt} \, \frac{\partial L}{\partial {\cal V}} = \frac{\partial L}{\partial x}   .
\end{equation}

They take the form obtained by writing a stationary action principle based on Eq. (\ref{app2}). Moreover, once the transformation $d/dt \rightarrow \dfr/dt$ is done, this form is nothing but the standard classical form. This result reinforces the identification of our tool with a ``quantum-covariant'' representation, since this Euler-Lagrange equation can be integrated in the form of a Schr\"odinger equation.


\end{document}